\documentclass{elsart5p}

\usepackage{graphicx}
\usepackage{amssymb}

\begin{document}

\begin{frontmatter}

\title{Ordered droplets in quantum magnets with long-range interactions}
%

\author{Thomas Vojta\corauthref{Name1}}
\ead{vojtat@umr.edu}
and
\author{Jos\'{e} A. Hoyos}

\address{Department of Physics, University of Missouri-Rolla,
             Rolla, Missouri, 65409, USA}

\corauth[Name1]{Corresponding author. Tel: (573) 341-4793 fax: (573) 341-4715}

\begin{abstract}
A defect coupling to the square of the order parameter in a nearly quantum-critical
magnet can nucleate an ordered droplet while the bulk system is in the paramagnetic
phase. We study the influence of long-range spatial interactions of the form
$r^{-(d+\sigma)}$ on the droplet formation. To this end, we solve a
Landau-Ginzburg-Wilson free energy in saddle point approximation. The long-range
interaction causes the droplet to develop an energetically unfavorable power-law tail. However,
for $\sigma>0$, the free energy contribution of this tail is subleading in the limit of
large droplets; and the droplet formation is controlled by the defect bulk. Thus, for
large defects, long-range interactions do not hinder the formation of droplets.
\end{abstract}


\end{frontmatter}


When a magnetic system is close to its critical point, a local defect can induce the
nucleation of a magnetic droplet in the nonordered bulk. In disordered systems, such
droplets are responsible for Griffiths singularities \cite{Griffiths69} and other rare
region effects \cite{Vojta06}. At zero-temperature quantum phase transitions, droplets
can lead to strong power-law quantum Griffiths singularities \cite{ThillHuse95} or even
to the smearing of the transition \cite{Vojta03}. A particularly interesting problem
arises in metallic quantum ferromagnets because the coupling between magnetic modes and
gapless fermions generates an effective long-range (power-law) spatial interaction
between the magnetic fluctuations \cite{VBNK96,BelitzKirkpatrickVojta05}. This
interaction takes the form $r^{-(2d-1)}$ for clean electrons and $r^{-(2d-2)}$ for
diffusive electrons, where $d\ge 2$ is the spatial dimensionality. Understanding
disordered metallic quantum ferromagnets therefore requires analyzing the formation of
magnetic droplets in the presence of long-range interactions.

In this paper, we consider this problem within a $d$-dimensional quantum
Landau-Ginzburg-Wilson field theory with a general long-range attractive $1/r^{d+\sigma}$
interaction.
  The action reads $S=S_{\rm st} + S_{\rm dyn}$ with the static part given by
\begin{eqnarray}
S_{\rm st} &=& \int d\tau\int d\mathbf{x}
d\mathbf{y}\varphi\left(\mathbf{x},\tau\right)
             \Gamma\left(\mathbf{x},\mathbf{y}\right)\varphi\left(\mathbf{y},\tau\right)
             \nonumber\\
         &+&\frac{u}{2}\int d\tau d\mathbf{x}\varphi^{4}\left(\mathbf{x},\tau\right)~.
\label{eq:S_stat}
\end{eqnarray}
$\varphi(\mathbf{x},\tau)$ is the order parameter at position $\mathbf{x}$ and imaginary time $\tau$.
The two-point vertex, $\Gamma\left(\mathbf{x},\mathbf{y}\right)= \Gamma_{\rm
NI}(\mathbf{x})\delta\left(\mathbf{x}-\mathbf{y}\right) +\Gamma_{\rm
I}\left(\mathbf{x},\mathbf{y}\right)$, contains a non-interacting part and the attractive
long-range interaction,
\begin{equation}
\Gamma_{\rm I}\left(\mathbf{x},\mathbf{y}\right)=
     -\gamma\left[\xi_{0}^{2}+\left|\mathbf{x}-\mathbf{y}\right|^{2}\right]^{-(d+\sigma)/{2}}.
     \label{eq:I_kernel}
\end{equation}
Here, $\xi_0$ is a microscopic cutoff length scale. To ensure a proper thermodynamic limit, the range parameter
 $\sigma$ must be positive. The noninteracting part of the vertex reads
\begin{equation}
\Gamma_{\rm NI}\left(\mathbf{x}\right)=t_{0}+\delta t\left(\mathbf{x}\right)+\Gamma_{0},
\label{eq:NI_kernel}
\end{equation}
where $t_{0}$ is the bulk distance from criticality, and the constant $\Gamma_{0}$  cancels the
$\mathbf{q}=0$ Fourier component of the interaction. $\delta t\left(\mathbf{x}\right)$
is the defect potential. For definiteness we consider a single spherical defect of radius $a$
at the origin with potential $\delta t (\mathbf{x}) = -V$ for $|\mathbf{x}|<a$, and  $\delta t
(\mathbf{x})=0$
otherwise. We are interested in defects that favor the ordered phase, i.e., $V>0$.

The \emph{existence} of locally ordered droplets at the defect can be studied within
saddle-point approximation. Since time-independent saddle-point solutions give the
lowest free energies, the static action (\ref{eq:S_stat}) is sufficient for this purpose.
To study the droplet quantum \emph{dynamics} one has to specify the dynamical action
$S_{\rm dyn}$. This has been reported in Ref.\ \cite{HoyosVojta07}.


To find saddle-point solutions we set $\varphi(\mathbf{x},\tau)=\phi(\mathbf{x})$
and minimize the total action with respect to $\phi(\mathbf{x})$. This leads to the saddle-point equation
\begin{equation}
\Gamma_{\rm NI}(\mathbf{x})\phi\left(\mathbf{x}\right)+u\phi^{3}
\left(\mathbf{x}\right) = \! \int \! \frac{\gamma\phi\left(\mathbf{y}\right)d\mathbf{y}}
{\left[\xi_{0}^{2}+\left|\mathbf{x}-\mathbf{y}\right|^{2}\right]^{(d+\sigma)/2}}.
\label{eq:Saddle-point_stat}
\end{equation}
We have not managed to solve this nonlinear integral equation in closed form.
In Ref.\ \cite{HoyosVojta07}, we have performed an asymptotic large-distance analysis
based on the ansatz
\begin{equation}
\phi\left(\mathbf{x}\right)= \left\{ \begin{array}{lr} \phi_{0} & \quad
(|\mathbf{x}|<a)\\ C/|\mathbf{x}|^{d+\sigma} & (|\mathbf{x}|> a)\end{array}\right. ~,
\label{eq:ansatz_phi}
\end{equation}
suggested by Griffiths' theorem \cite{Griffiths67}. The parameters $\phi_0$ and $C$
follow from solving the linearized saddle-point equation and minimizing the action.
For large defects ($a\gg \xi_0$), they read
\begin{equation}
C=\Omega_d\phi_{0}\gamma\, a^{d}/ (dt_{0}) ~, \quad \phi_{0}=\sqrt{(V-t_{0})/{u}}~.
\label{eq:parameters}
\end{equation}
As a result, the saddle point action takes the form
\begin{equation}
S_{\rm SP}=\frac{\Omega_d}{d}\phi_{0}^{2} a^{d}\left(t_{0}-V+\frac{u}{2}\phi_{0}^{2}\right)+{\cal O}\left(a^{d-1},a^{d-2\sigma}\right)
\label{eq:SP-action}
\end{equation}
to leading order in the defect size. This result, which is identical to the case of
short-range interactions \cite{CastroNetoJones00,MillisMorrSchmalian02}, implies that the droplet
formation is dominated by the defect core. The power-law droplet tail thus does not hinder
the droplet formation.

Here, we compare the results of the asymptotic analysis \cite{HoyosVojta07} with a
numerically exact solution of the saddle-point equation. We focus on three space
dimensions and the cases $\sigma=2$ and 3 not considered in \cite{HoyosVojta07}.
The numerical procedure is as follows: For our spherical defect, the angular integrations
on the right-rand side of (\ref{eq:Saddle-point_stat}) can be carried out analytically.
The resulting one-dimensional equation  is discretized (using $10^5$
sites)
and solved iteratively.

\begin{figure}
\quad\includegraphics[clip=true,width=6.5cm]{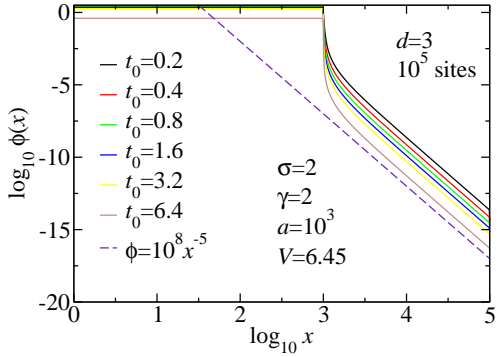}
\caption{Local order
         parameter $\phi$ of a three-dimensional droplet with $\sigma=2$
          as a function of distance $x$ from the defect center for different distances
          $t_0=0.2$ to 6.4 from bulk criticality (from top to bottom).}
\label{fig:1}
\end{figure}

\begin{figure}
\quad\includegraphics[clip=true,width=6.5cm]{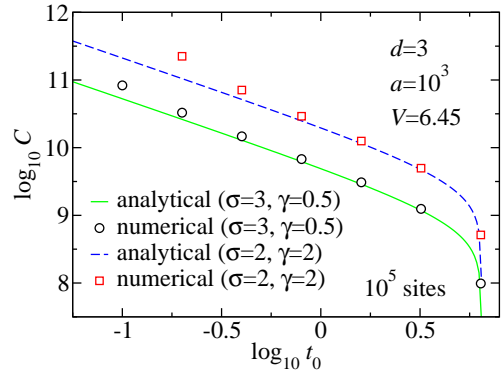} \caption{Amplitude $C$ of
the power-law decay of the droplet tail for the system shown in Fig.\ \ref{fig:1} and the
analogous results for $\sigma=3$. The lines are the prediction (\ref{eq:parameters}).}
\label{fig:2}
\end{figure}

Figure \ref{fig:1} shows saddle-point solutions for $d=3,\sigma=2$ and several
 $t_0$.
In agreement with (\ref{eq:ansatz_phi}), the droplet tail falls off like
$|\mathbf{x}|^{-5}$. The solution for $\sigma=3$ is completely analogous, with the tail
dropping off as $|\mathbf{x}|^{-6}$, as predicted. The amplitudes of these power-law
decays are analyzed in Fig.\ \ref{fig:2}.
As predicted in (\ref{eq:parameters}), for small $t_0$, $C$ behaves like $1/t_0$
(the small deviations at very low $t_0$ are finite-size effects).

In summary, we have studied how a magnetic droplet nucleates at a defect coupling to the
square of the order parameter in an quantum magnet with long-range interactions of the
form $r^{-(d+\sigma)}$ with $\sigma>0$. The droplet magnetization develops a long
power-law tail, i.e., at large distances $r$ from the defect, it decays like
$r^{-(d+\sigma)}$ in agreement with Griffiths' theorem \cite{Griffiths67}. However, the
droplet free energy is dominated by the core contribution while the tail contribution is
subleading in the limit of large defects. Therefore, the droplet formation is analogous
to the case of short-range interactions.

These results are potentially important for quantum phase transitions in disordered
metallic ferromagnets such as Ni$_{1-x}$Pd$_{x}$ \cite{Nicklas},
URu$_{2-x}$Re$_{x}$Si$_2$ \cite{Bauer}, or Fe$_{1-x}$Co$_x$S$_2$ \cite{DiTusa}. Our
calculations show that in these systems, locally ordered droplets can form on rare
(strongly coupled) spatial regions. We predict that the slow dynamics of these magnetic
droplets leads to strong power-law quantum Griffiths effects close to the transition,
(for Heisenberg symmetry) \cite{VojtaSchmalian05} or a destruction of the phase
transition by smearing (for Ising symmetry) \cite{Vojta03} just like for systems with
short-range interactions \cite{Vojta06}, provided the droplet-droplet interactions are
negligible (see Ref.\ \cite{HoyosVojta07}).

This work was supported by the NSF under grant no. DMR-0339147 and by Research Corporation.


\begin{thebibliography}{99}

\bibitem{Griffiths69} R.B.\ Griffiths, Phys.\ Rev.\ Lett.\ {\bf 23} (1969), 17.

\bibitem{Vojta06} T.\ Vojta, J.\ Phys.\ A {\bf 39} (2006), R143.

\bibitem{ThillHuse95} M.\ Thill and D.A.\ Huse, Physica A {\bf 214} (1995) 321.

\bibitem{Vojta03} T.\ Vojta, Phys.\ Rev.\ Lett.\ {\bf 90} (2003) 107202.

\bibitem{VBNK96} T.\ Vojta, D.\ Belitz, R.\ Narayanan, and T.R.\ Kirkpatrick,
        Europhys.\ Lett.\, {\bf 36} (1996) 191.

\bibitem{BelitzKirkpatrickVojta05} D.\ Belitz, T.R.\ Kirkpatrick, and T.\ Vojta, Rev.\
        Mod.\  Phys.\ {\bf 77} (2005) 579.

\bibitem{HoyosVojta07} J.A.\ Hoyos and T.\ Vojta, Phys. Rev. B. {\bf 75},104418 (2007).

\bibitem{Griffiths67} R.B.\ Griffiths, J.\ Math.\ Phys.\ {\bf 8} (1967) 478.

\bibitem{CastroNetoJones00} A.H. Castro Neto and B.A. Jones, Phys. Rev. B {\bf 62} (2000)
     14975.

\bibitem{MillisMorrSchmalian02} A. Millis, D.K. Morr, and J. Schmalian, Phys. Rev. Lett
{\bf 87} (2002) 167202.

\bibitem{Nicklas} M. Nicklas et al., Phys. Rev. Lett. {\bf 82}, 4268 (1999).

\bibitem{Bauer} E. Bauer et al., Phys. Rev. Lett {\bf 94}, 046401 (2005).

\bibitem{DiTusa} J. DiTusa et al., cond-mat/0306541.

\bibitem{VojtaSchmalian05} T. Vojta and J. Schmalian, Phys. Rev. B {\bf 72}, 045438
(2005).
\end{thebibliography}
\end{document}